\documentclass[sigconf]{acmart}

\AtBeginDocument{%
  \providecommand\BibTeX{{%
    \normalfont B\kern-0.5em{\scshape i\kern-0.25em b}\kern-0.8em\TeX}}}

\setcopyright{acmcopyright}
\copyrightyear{2021}
\acmYear{2021}
\acmDOI{}

\acmConference[SEIS '22]{SEIS '22: Software Engineering in Society}{May 21--29, 2022}{Pittsburgh, PA}
\acmBooktitle{SEIS '21: Software Engineering in Society,
  May 21--29, 2018, Pittsburgh, PA}
\acmPrice{}
\acmISBN{}

\usepackage{graphicx}
\usepackage{xurl}
\usepackage{booktabs}
\usepackage{multibib}
\usepackage{natbib}
\usepackage{hhline}
\usepackage{xcolor}
\usepackage{graphicx}
\usepackage{colortbl}
\usepackage{multirow}
\usepackage{tabularx}

\newcommand{\MyBox}[1]{\vspace{3mm}\noindent\framebox[\columnwidth][c]{\parbox[b]{0.95\columnwidth}{ #1 }}\vspace{3mm}}

\copyrightyear{2022} 
\acmYear{2022} 
\setcopyright{acmcopyright}\acmConference[ICSE-SEIS'22]{Software Engineering in Society}{May 21--29, 2022}{Pittsburgh, PA, USA}
\acmBooktitle{Software Engineering in Society (ICSE-SEIS'22), May 21--29, 2022, Pittsburgh, PA, USA}
\acmPrice{15.00}
\acmDOI{10.1145/3510458.3513018}
\acmISBN{978-1-4503-9227-3/22/05}

\begin{document}

\title{An Empirical Investigation on the Challenges Faced by Women in the Software Industry: A Case Study}


\author{Bianca Trinkenreich}
\affiliation{%
  \institution{Northern of Arizona University}
  \city{Flagstaff, AZ}
  \country{USA}}
\email{bianca_trinkenreich@nau.edu}

\author{Ricardo Britto}
\affiliation{
  \institution{Ericsson \\ Blekinge Institute of Technology}
  \city{Kalrksona}
  \country{Sweden}}
\email{ricardo.britto@ericsson.com}

\author{Marco A. Gerosa}
\affiliation{
  \institution{Northern Arizona University}
  \city{Flagstaff, AZ}
  \country{USA}}
\email{marco.gerosa@nau.edu}

\author{Igor Steinmacher}
\affiliation{
  \institution{Univ. Tecnológica Federal do Paraná}
  \country{Brazil}}
\email{igorfs@utfpr.edu.br}


\begin{abstract}
\textbf{Context:} Addressing women's under-representation in the software industry, a widely recognized concern, requires attracting as well as retaining more women. Hearing from women practitioners, particularly those positioned in multi-cultural settings, about their challenges and and adopting their lived experienced solutions can support the design of programs to resolve the under-representation issue. \textbf{Goal:} We investigated the challenges women face in global software development teams, particularly what motivates women to leave their company; how those challenges might break down according to demographics; and strategies to mitigate the identified challenges. \textbf{Method:} To achieve this goal, we conducted an exploratory case study in Ericsson, a global technology company. We surveyed 94 women and employed mixed-methods to analyze the data. \textbf{Results:} Our findings reveal that women face socio-cultural challenges, including work-life balance issues, benevolent and hostile sexism, lack of recognition and peer parity, impostor syndrome, glass ceiling bias effects, the prove-it-again phenomenon, and the maternal wall. The participants of our research provided different suggestions to address/mitigate the reported challenges, including sabbatical policies, flexibility of location and time, parenthood support, soft skills training for managers, equality of payment and opportunities between genders, mentoring and role models to support career growth, directives to hire more women, inclusive groups and events, women's empowerment, and recognition for women's success. The framework of challenges and suggestions can inspire further initiatives both in academia and industry to onboard and retain women.

\end{abstract}

%
%


\maketitle

\section*{Lay Abstract} 
 Women represent less than 24\% of employees in software development industry and experience various types of prejudice and bias. Even in companies that care about Diversity \& Inclusion, ``untying the mooring ropes" of socio-cultural problems is hard.
Hearing from women, especially those working in a multi-cultural organization, about their challenges and adopting their suggestions can be vital to design programs and resolve the under-representation issue. In this work we work closely with a large software development organization which invests and believes in diversity and inclusion. We listened to women and the challenges they face in global software development teams of this company and what these women suggest reduce the problems and increase retention. Our research showed that women face work-life balance issues and encounter invisible barriers that prevent them from rising to top positions. They also suffer micro-aggression and sexism, need to show competence constantly, be supervised in essential tasks, and receive less work after becoming mothers. Moreover, women miss having more female colleagues, lack self-confidence and recognition. The women from the company suggested sabbatical policies, the flexibility of location and time, parenthood support, soft skills training for managers, equality of opportunities, role models to support career growth, directives to hire more women, support groups, and more interaction between women, inclusive groups and events, women's empowerment by publishing their success stories in media and recognizing their achievements. Our results had been shared with the company Human Resources department and management and they  considered the diagnosis helpful and will work on actions to mitigate the challenges that women still perceive.

\section{Introduction}
\label{sec:introduction}


Diverse software teams are more likely to understand user needs, contributing to a better alignment between the delivered software and its intended customers~\citep{muller1993participatory}. Diversity further positively affects productivity by bringing together different perspectives~\cite{vasilescu2015gender} and fosters innovation and problem-solving capacity, leading to a healthier work environment~\citep{earley2000creating}. However, women are still underrepresented in the software industry~\cite{hyrynsalmi2019underrepresentation}. Reducing the gender gap in the software industry requires not only attracting, but also \textit{retaining women}. 

Women\footnote{\label{foot1}In this study, we use the term ``gender'' as a socially constructed concept~\cite{butler1999gender}, where gender identification, display, and performance might or might not align with a person's sex assigned at birth. To reflect this social concept of gender, we use the term ``women'' and ``men'' as a shorthand for people who self-identify as such.} often face socio-cultural challenges in the software industry and could decide to leave their jobs (or even the software industry) if diversity is not a priority~\cite{Lee.Carver:2019}. Gender-related incidents can be so severe that they motivate women to leave a project or their jobs~\cite{paul2019expressions,vasilescu2015perceptions}. Kuechler et al.~\cite{kuechler2012gender} suggest that women drop out because their jobs are not aligned to their motivations or due to the unappealing and hostile social dynamics in their daily work. Understanding the reasons behind the decision to step out of a project or role can help create strategies to increase retention. 

Challenges faced by women have been largely investigated in the free and open source software (F/OSS) development context (e.g., \cite{Lee.Carver:2019, balali2018, powell2010gender, canedo2020work, imtiaz2018, singh2019women2,calvo2021visible,Nafus2012NMS, paul2019expressions, qiu2019signals, terrell2017gender, vedres2019gendered, vasilescu2015perceptions}). Although there are a few studies focusing on software companies \cite{Canedo2021SBES, wolff2020prevents, musungwini2020challenges, orser2012perceived}, the results are still preliminary and more studies are needed to contribute to theory building. In this context, theory is built by aggregating results from case studies and related qualitative research by making comparisons, looking for similarities and differences within the collected data, and by examining future questions~\cite{lawrence2014social}.

In this paper, we contribute to this scientific body of knowledge by reporting a case study in a multi-cultural global software development organization from a large company, namely Ericsson, which is a global private company that has more than 100,000 employees around the world, and is one of the leading providers of Information and Communication Technology (ICT). One of the Diversity \& Inclusion business goals of the company is to reach at least 25\% proportion of women in every suborganization of the company. However, the software development organization still struggles to attract and retain women and has partnered with external researchers to understand the challenges and strategies from the point of view of the women themselves. As the phenomenon under study is complex because it involves different perspectives and disciplines, it is critical to consider how women perceive and give meaning to their social reality.

Therefore, our goal is to identify the challenges that women face in the company, according to them, and investigate the measures that women recommend to mitigate the identified challenges. To achieve our goal, we defined the following research questions:\\
\textit{RQ1: What challenges do women face in software teams?\\
RQ2: What are possible actions to mitigate the identified challenges, from the women's perspectives?}

To answer our research questions, we collected data through an online questionnaire that included questions about the challenges that are currently being faced, reasons women would decide to leave, and suggestions to increase women's participation. The questionnaire was answered by 94 women from a software development suborganization of Ericsson. The novelty of the work includes considering a multi-cultural global software development organization from a large IT company, breaking down the challenges according to demographics, identifying challenges that push women out of the company, and connecting challenges to potential strategies to attract and retain women from the point of view of the women themselves, who are on the front lines of the problem.

We introduce our research design in Section~\ref{sec:research_design} and the study results in Section~\ref{sec:results}. Section~\ref{sec:discussion} discusses the results and implications of our results, followed by related work in Section~\ref{sec:related_work}, limitations, and conclusions in Sections~\ref{sec:threats} and \ref{sec:conclusion}.

\section{Research Design}
\label{sec:research_design}
To answer our research questions, we conducted an exploratory case study~\cite{runeson2012case} via a questionnaire administered to women from one of the software development suborganizations of Ericsson. In this section, we describe the case and the phases of planning, data collection, and analysis.

\subsection{The Case and Unit of Analysis}
\label{sec:case_description}

The case and unit of analysis is one of Ericsson's software development organizations. Ericsson is a global and large company that develops telecommunications-related products. It has more than 100,000 employees who are geographically distributed in several countries, including India, Sweden, Canada, USA, Poland, Brazil, and Germany. 


To diagnose how the company is doing in terms of gender diversity, every year, Ericsson\footnote{\url{http://www.ericsson.com}} publishes a diversity report with the percentage of employees who identify themselves as women\footnote{\label{Annual Diversity Report}\url{https://www.ericsson.com/en/investors/financial-reports/annual-reports}}. The company has a goal to achieve at least 25\% of women as employees in all suborganizations. This percentage was achieved for the whole company in 2020, but not in a consistent way in every suborganization. One of the software development organizations still dominated by men (the name of which is omitted for confidentially reasons) decided to conduct a systematic research to understand the viewpoint of the women who currently work at the company, in order to plan informed and bottom-up actions to mitigate the reported challenges. The company is interested in understanding the challenges women report as well as their suggestions to improve the current situation, as a contemporary phenomenon within its real-life context~\cite{yin2009case}. This case study was collaboratively conducted by one researcher from the company and three outsiders.


\subsection{Data Collection}
\label{sec:method:data_collection}

We administered an online questionnaire~\footnote{The research protocol was approved by the university institutional review board (IRB).} using the Qualtrics tool\footnote{\url{http://www.qualtrics.com}} to employees from the software development suborganization who self-identify as women. We opted for a questionnaire instead of interviews to increase participation and coverage of the research. The questionnaire was designed to understand the state of the problem, asking open questions about the challenges faced, the reasons that drove women they know to leave the company, and their suggestions to retain more women in the company. The demographic questions were the last part of the questionnaire, and were chosen to help us investigate possible relations between groups. Although family is universal to all genders, women culturally face a greater pressure to balance work and family~\cite{sturges2004working}. Considering women as the target population of the present study, besides country, age, and experience, we included demographics questions to understand possible intersections of the challenges and family status. All questions were optional to increase the response rate by making respondents more comfortable~\cite{punter2003conducting}. After proofreading and testing in multiple browsers and devices, we invited three participants to pilot the questionnaire so we could collect feedback and measure the time to answer. No modification of the questionnaire was necessary, and we discarded these initial answers.

The managers of the company software teams sent emails to the women on their teams with the questionnaire link. Every week, the first author of this paper followed up with the company's manager to check the number of answers and identify sites that needed additional encouragement. 

The questionnaire was open for answers between June 11 and July 20, 2021. Our questionnaire received 94 non-blank answers. We report the demographics of the respondents in Table~\ref{tab:demographics}. Most of the respondents were residents of India (65.6\%), less than 35 years old (50.0\%), married (62.8\%), living with children (52.1\%), having more than 10 years of experience in the software industry (45.7\%) and less than 5 years working in the company (35.1\%). A balanced proportion of respondents have and do not have children living with them. The sample mirrors the company's numbers. In 2020, the prevalent nationality (55\%) of the employees from the studied suborganization was from India. Although we do not have data about the employees' age of each suborganization, 31\% of the overall number of employees age less than 35 years old. The other demographics are not measured by the company's annual report.

\begin{table}[tbh]
\centering
\caption{Personal characteristics of the respondents (n=94)}
\label{tab:demographics}
\begin{tabular}{l|r|r}
\hline
\toprule
 \textbf{Demographics} & \textbf{\#}  & \textbf{\%}  \\
 \hline

\toprule
Experience: $\leq$ 5 years in software industry                     & 12  & 12.8\% \\
Experience: $>$ 5 \& $<$ 10 years in software industry & 21  & 22.3\% \\
Experience: $\geq$ 10 years in software industry     & 43  & 45.7\% \\
Did not inform & 18 & 19.1\% \\
\toprule
Tenure: $\leq$ 5 years in Ericsson                     & 33 & 35.1\% \\
Tenure: $>$ 5 \& $<$ 10 years in Ericsson & 26  & 27.7\% \\
Tenure: $\geq$ 10 years in Ericsson     & 24  & 25.5\% \\
Did not inform & 11 & 11.7\% \\
\toprule
Age: Less than 35                                      & 47  & 50.0\% \\
Age: 35 to 44                                        & 27  & 28.7\% \\
Age: 45 to 54                                        & 10  & 10.6\% \\
Did not inform                                   & 10   & 10.6\% \\
\toprule
Country: India                              & 61  & 64.9\% \\
Country: Brazil                                 & 11 & 11.7\% \\
Country: Canada                                 & 8   & 8.5\%  \\
Country: Others                                & 5  & 5.3\% \\
Did not inform                                   & 9   & 9.6\% \\
\toprule
Marital Status: Married or domestic partnership     & 59  & 62.8\% \\
Marital Status: Single or divorced         & 24  & 25.5\% \\
Did not inform      & 11  & 11.7\% \\
\toprule
Have children living with: Yes         & 49  & 52.1\% \\
Have children living with: No         & 36  & 38.3\% \\
Did not inform      & 9  & 9.6\% \\
\bottomrule
\end{tabular}
\end{table}

We filtered our data to consider only valid responses. We manually inspected the open text questions, looking for blank answers. Instead of removing the entire response when one of the questions was blank or did not report challenges or suggestions, we separated data into two datasets of 94 answers each: one dataset for challenges and one dataset for suggestions. Then, we named each respondent from S1 to S94, removed blanks for each dataset (3 removed from challenges and 15 for suggestions). Next, we removed the answers that were not informing any challenge or suggestion (27 removed from challenges who reported that they do not face any challenge and 9 from the suggestions who reported having no proposed solutions). The final step was to check potential duplicate participation, even though the tool used in our investigation (Qualtrics) has mechanisms to prevent multiples responses from the same participant. The final challenges' dataset had 64 answers, while the suggestions' dataset had 70 answers. 

\subsection{Data Analysis} 
\label{sec:method:data_analysis}

To answer both RQ1 and RQ2, we analyzed the responses to the open questions about challenges, reasons for leaving, and suggestions to increase women's participation in the company. The first author qualitatively analyzed the answers for the open questions by inductively applying open coding\cite{miles1994qualitative} to organize what participants reported. We then organized our categories following concepts from existent theories, such as sexism (hostile and benevolent)~\cite{glick1996ambivalent}, impostor syndrome~\cite{clance1978imposter}, maternal wall~\cite{williams2007evolution}, prove-it again~\cite{biernat1997gender}, glass ceiling~\cite{jackson2001women,sharma2014glass}, work-life balance issues~\cite{guest2002perspectives}, lack of peer parity~\cite{ford2017someone}, and lack of recognition~\cite{arlow1955motivation}. We built post-formed codes, having three of the authors conducting card sorting sessions~\cite{Spencer2009}, including discussing the codes and categorization until reaching consensus about the the codes and the corresponding literature. 

After completing the qualitative analysis, we checked the distribution of answers categorized in each challenge. From the 64 women who reported they face some type(s) of challenges, 34 reported challenges related to only one category, 25 to two different categories, 3 to three categories, and 2 women reported challenges related to four categories. We also used descriptive statistics to summarize the responses and their association with the demographics data~\cite{wohlin2015towards}.


To analyze how the challenges differ according to individual characteristics, we segmented our sample based on \emph{experience in software industry} (experienced: $\geq$ 10 years of experience vs. less experienced: $\leq$ 10 years of experience), \emph{tenure or years in company}(more years in company: $\geq$ 10 years in company vs. less years in company: $\leq$ 10 years in company), \emph{age} (older: $\geq$ median, 35 years old vs. younger: $<$ median, 35 years old), \emph{married or not}, and \emph{lives, or not, with children}.

See supplementary material\footnote{https://figshare.com/s/d1c3bd386083fa55104a} for additional details, including the answers to the demographics, the open questions and the qualitative analysis codes.

Next, we calculated the odds ratio for each challenge and demographic information. We interpreted the results as follows: 
\begin{itemize}
    \item if \textbf{Odds Ratio $=$ 1}, both groups are equally distributed for the reported challenge. 
    \item if \textbf{Odds Ratio $>$ 1}, the likelihood for the reported challenge is higher for the first group (in our case: experienced, older than 35, married and living with children). 
    \item if \textbf{Odds Ratio $<$ 1}, the likelihood for the reported challenge is higher for the second group (in our case: novices, younger than 35 years, single or divorced, without children). 
\end{itemize}

\section{Study Results}
\label{sec:results}

In this section, we present the results of our investigation, which are grouped by research question.

\subsection{RQ1: What challenges do women face in software development teams?}
\label{sec:results:rq1}

\begin{figure*}[!bth]
\centering
\includegraphics[width=1\textwidth]{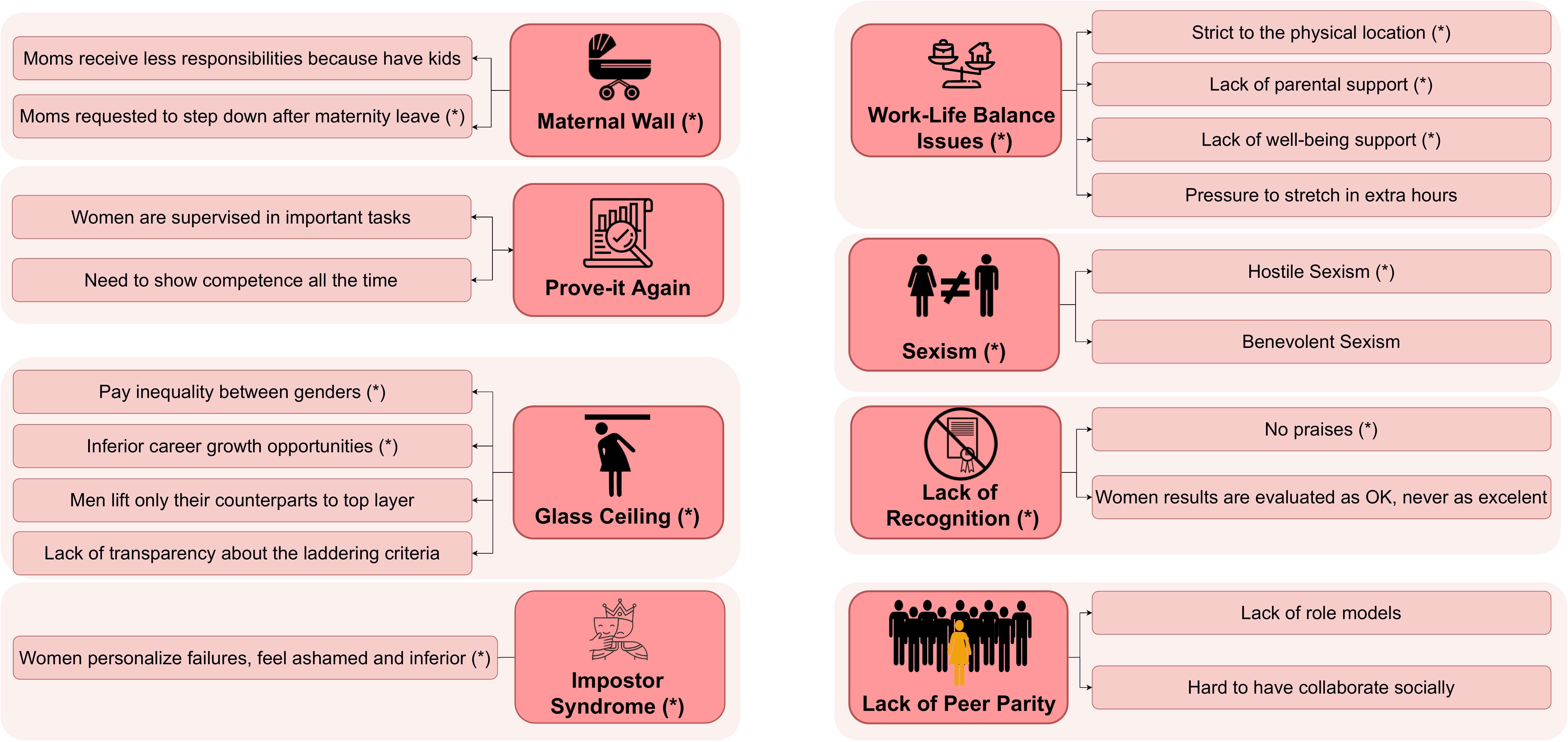}
\caption{The challenges reported by women who participated in our study. Those marked with an asterisk (*) were reported as a challenge that ultimately can lead women to leave the company.}
\label{fig:challenges}
\end{figure*}

We found eight challenge categories, as presented in Fig.~\ref{fig:challenges}. We marked with an asterisk (*) the challenges reported by at least one person as a reason to leave the company. Table \ref{tab:codes_challenges} presents the number of participants whose responses fit in each category. In the following, we present more details about our findings organized by challenge category.

\begin{table*}[!tbh]
\centering
\caption{Representative examples of answers to the challenges' open question, number and percentage of women whose answer was coded for each category. In parenthesis the number of women who reported that challenge as a reason to leave.}
\label{tab:codes_challenges}
\begin{tabular}{c|l|r|r}
\hline
\toprule
 \textbf{Challenge} & \textbf{Representative examples} & \textbf{\#}  & \textbf{\% (n=64)}  \\
 \hline

\toprule
\begin{tabular}[c]{@{}c@{}}Work-Life\\Balance Issues\end{tabular} & \begin{tabular}[l]{@{}l@{}}\textit{``To be successful, any professional is expected to work super-hard and go far and beyond,}\\ \textit{working overtime, constantly learn new things and this takes a lot of energy, but after my}\\ \textit{working hours end, my home, my second and more important work starts..."} (S40)\end{tabular} & 35 (25) & 54.7\% \\ \hline
Sexism & \begin{tabular}[l]{@{}l@{}}\textit{``In technical discussions, people feel that women cannot do it better so they make} \\\textit{comments which are makes you uncomfortable"} (S23)\end{tabular}& 14 (3) & 21.9\% \\ \hline
Lack of Recognition & \textit{``Lack of praise"} (S38). \textit{Not been recognized by the job"} (S63)& 2 (2) & 3.1\% \\ \hline
Lack of Peer Parity & \begin{tabular}[l]{@{}l@{}}\textit{``Working in a place where there is a dominant number of male colleagues, inclusion in all}\\\textit{discussions might not be uniform across" (S43)}\end{tabular}& 9 & 14.1\% \\ \hline
Impostor Syndrome & \begin{tabular}[l]{@{}l@{}}\textit{``When it comes to failure, the failures are easily owned and personalised by a women}\\ \textit{and they on resign their own compared to men"} (S57)\end{tabular} & 2 (1) & 3.1\% \\ \hline
Glass Ceiling & \begin{tabular}[l]{@{}l@{}}\textit{``Glass roof, only a few women as leaders"} (S38). \textit{``When women try to achieve more things}\\\textit{they are called ambitious in a negative way. Whereas men are expected to." (S11)}\end{tabular}& 26 (19) & 40.6\% \\ \hline
Prove-it Again & \begin{tabular}[l]{@{}l@{}}\textit{``As a woman I need to work harder to achieve the same as a man. I need to show}\\ \textit{competence and to be 100\% right all the time."} (S25)\end{tabular} & 9 & 14.1\% \\ \hline
Maternal Wall & \begin{tabular}[l]{@{}l@{}}\textit{``Some people treat you differently because you just had a kid, giving you less work [..]}\\ \textit{and framing you inside a box"} (S79).\end{tabular} & 4 (1) & 6.3\% \\
\bottomrule
\end{tabular}

\resizebox{0.75\textwidth}{!}{%
\tiny
\begin{tabular}{r}
The total per challenge is not the sum of the respondents since the participants often provided an answer that was categorized into \textbf{more than one challenge}.
\end{tabular}
}
\vspace{-0.2cm}

\end{table*}

\subsubsection{Work-Life Balance Issues}

Our participants reported that before the COVID-19 pandemic the company was \textsc{strict to the physical location} and many women left due to facing the ``trailing spouse" effect~\cite{harvey1998dual,branden2014gender}, moving to another city or country when their spouse was relocated. 

After the COVID-19 pandemic and suddenly working from home, women reported facing \textsc{pressure to work extra hours}, \textit{``having no boundary set for working hours''} (S24), and having to either attend meetings in different time zones or to learn new knowledge for the work. They also mentioned the consequences for not giving in to this pressure: they would be excluded from decisions that are made during the meetings and are perceived by others as \textit{``lacking in teamwork''} (S30). 

When working extra and long hours, women feel stressed and have trouble disconnecting from work, \textit{``impacting other household chores and having hardly any time left to bring some peace to the mind''} (S49). The \textsc{lack of well-being support} causes high levels of stress, which would be one of the reasons to quit. 

Besides the inflexible location, our participants also mentioned that before the COVID-19 they lacked flexible work hours (which improved during the pandemic) and  paid sick leave (specific for one location, for which the local laws do not cover this). These points are really important for those who have  parenting and caretaking responsibilities, though. Lack of daycare in the office was associated with \textsc{lack of parental support}, which can cause women to leave the company.

\subsubsection{Sexism}
The ambivalent sexism theory~\cite{glick1996ambivalent} defines sexism as a multidimensional construct that includes two sets of attitudes: hostile and benevolent. Sexism has typically been conceptualized as a reflection of hostility toward women~\cite{glick1996ambivalent}. \textsc{Hostile sexism} is related to the classic definition of prejudice~\cite{allport1954nature}. 

Our participants from our investigation reported microaggressions in which their \textit{``voices are suppressed per opposite gender''} (S19). Besides not being heard during technical discussions, women receive various diminishing comments, such as that women cannot bring the same results as men and that \textit{``women come to work only for time pass or are not brilliant enough"} (S85). Exposure to such diminishing comments can be a reason to leave.  

\textsc{Benevolent sexism} represents the subjectively positive feelings toward a gender that often bring some sexist antipathy~\cite{glick1996ambivalent}. The participants reported being \textit{``pampered, never been given a hard/straight feedback''} (S57) and being included in initiatives only because they are women, not because of their skills and capacity.

\subsubsection{Lack of Recognition}
Feeling valued or appreciated is part of Maslow's hierarchy of human needs~\cite{arlow1955motivation}. The participants mentioned \textit{``Not been recognized by her job''} (S63) and that \textsc{\textit{women's results are usually evaluated as OK, never as excellent}} (S57), even when accomplishing exceptional work. \textsc{No praises} from managers was considered one of the reasons to leave.

\subsubsection{Lack of Peer Parity} 
Being surrounded by similar individuals to which to compare oneself, or identifying with at least one other peer in the team, is known as peer parity~\cite{ford2017someone}. 

The participants mentioned an \textit{``[im]balance in men:women ratio''} (S37) and two consequences: (i) impact on their social capital, as they considered it \textsc{\textit{``hard to collaborate socially''}} (S55), \textit{``[be]cause men socialize in a different way than women do''} (S12); and (ii) impact on developing their self-confidence due to \textsc{lack of role models}, as they \textit{``lack [] strong women leaders as mentors''} (S14).

\subsubsection{Impostor Syndrome} 
Impostor Syndrome (also known as impostor phenomenon, fraud syndrome, perceived fraudulence, or impostor experience), describes an experience of individuals who, despite their objective successes, struggle to internalize their accomplishments, feel persistent self-doubt, and being exposed as a fraud or impostor~\cite{clance1978imposter}. 

Our participants mentioned as a challenge and reason to leave situations in which \textsc{women personalize failures} \textsc{\textit{and ``feel ashamed and inferior}} \textit{more than men and they tend to escape it by leaving [the company], but always masked as personal reasons''} (S6).

\subsubsection{Glass Ceiling} describes a corporate world phenomenon in which minorities' access to the top-management positions is blocked by tradition or culture~\cite{jackson2001women}, as an invisible structural barrier that prevent minorities from career advancement~\cite{sharma2014glass}. 

Two reasons to leave reported by the participants included the perception of \textsc{pay inequality between genders} and \textsc{inferior career growth opportunities} for women. For the former, S30 stated that \textit{``women employees are paid less compared to male counterparts''}, while for the last, S69 mentioned that she \textit{``reach[ed] a stage where [they] have nowhere to [climb next] in the ladder''} (S69). Still, women reported that they feel that they work harder to achieve the same positions as men, indicating a possible \textsc{lack of transparency about the ladder criteria}, as their ambition is discouraged, while \textit{``corporate politics are played by men''} (S21) and \textsc{men lift only their counterparts to top layer}.

\subsubsection{Prove-it Again} effect is a bias that occurs when a member of a group that does not align with stereotypes is measured at a stricter standard than those who do align with the stereotypes and, consequently, has to constantly provide more evidence to demonstrate competence~\cite{biernat1997gender}. 

The participants mentioned that women \textsc{need to show competence all the time}: \textit{``put extra effort to be heard when there is competition between men''} (S84) and having \textit{`` no room to slip [up]"} (S41). Lastly, women feel they need to prove themselves \textsc{when receiving an important task, [as] they are supervised} by another person to guarantee they do it correctly (S65).

\subsubsection{Maternal Wall} describes the experience of mothers whose coworkers perceive and judge them as having made one of two choices: either they continue to work and neglect their family, making the mother less likable, or the mother prioritizes family over work, making them less reliable in the workplace~\cite{williams2007evolution}. Our participants reported that women who are mothers \textsc{receive [fewer] responsibilities because they have kids}, as they are believed to not be able to handle much work. One of them reported \textit{``surprising colleagues that [they] are able to handle it all''} (S7). In addition, one of the reasons that cause women to leave is that \textit{``when returning from maternity leave, not enough support is provided or generally \textsc{the woman is asked to step down from the role}''} (S38).

\begin{table*}[bht]
\centering
\caption{Odds ratios per personal characteristic}
\label{tab:odds}
\begin{tabular}{l|l|l|l|l|l}
\toprule
& \textbf{\begin{tabular}[x]{@{}l@{}}More Experienced \\ vs. Less Experienced\end{tabular}} & \textbf{\begin{tabular}[x]{@{}l@{}}More Years in Company \\ vs. Less Years in Company\end{tabular}} & \textbf{
\begin{tabular}[x]{@{}l@{}}Older vs.\\ Younger\end{tabular}}  & 
\textbf{\begin{tabular}[x]{@{}l@{}}Married\\vs. Single\end{tabular}}
& \textbf{
\begin{tabular}[x]{@{}l@{}}Child vs.\\no child\end{tabular}} \\
\hline  
\midrule
Work-Life Balance Issues       & \cellcolor{gray!25}3.17** & 1.09 & 2.51 & \cellcolor{gray!25}5.08** & 3.29 \\
Glass Ceiling     & 0.48 & 1.29 & 0.38 & 0.76 & 1.46  \\
Sexism       & 0.46 & 0.34 & 0.47  & 0.37 & 2.13 \\
\bottomrule
\multicolumn{5}{l}{\footnotesize Significance codes: * $p<0.10$, ** $p<0.05$ *** $p<0.01$.}\\
\multicolumn{6}{l}{\scriptsize Note: Odds ratio $>$ 1 means that the first segment has greater chances of reporting the challenge than the second. Ratio $<$ 1 means the opposite. The challenges were coded from the open question.}
\end{tabular}
\end{table*}

\subsubsection{A segmented look at the challenges perceived by women}
\label{sec:segmented_analysis_challenges}

\begin{figure*}[!bht]
     \centering
     \includegraphics[width=1\textwidth]{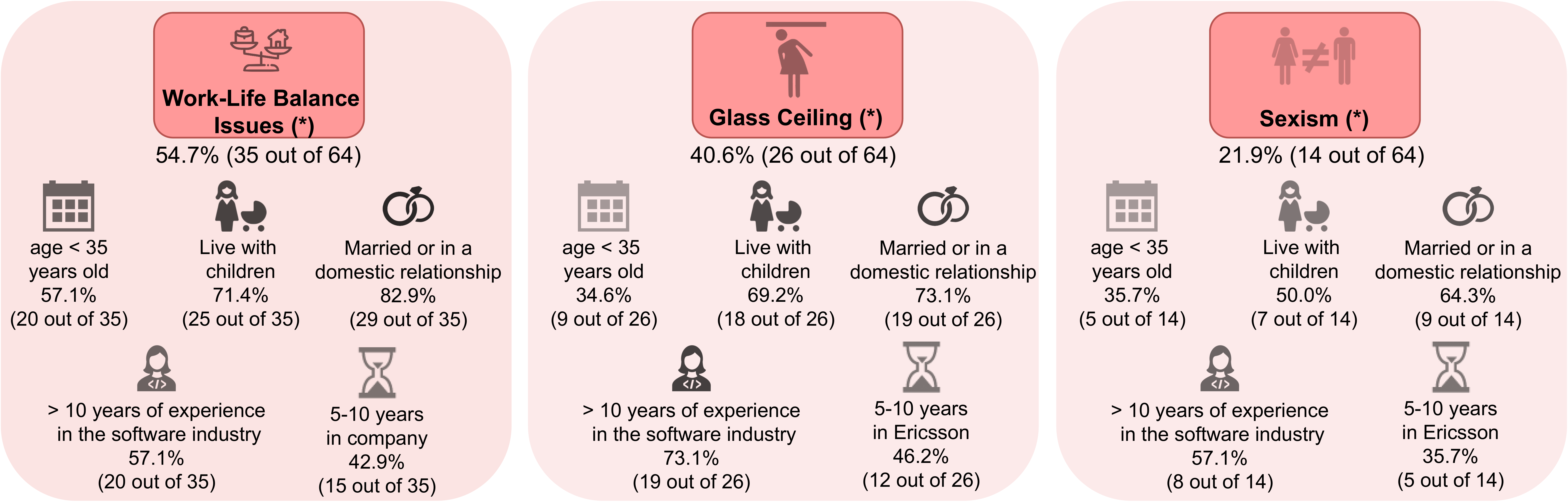}
     \vspace{-0.4cm}
     \caption{Subgroup analysis of the top three categories of challenges. The opacity of the icons represents the percentage of each category of challenge. Darker means a higher and lighter a lower percentage. The percentage below each challenge represents the number of respondents who reported that challenge. Some respondents provided answers about challenges that accounted for more than one category.}
     \label{fig:segmented_analysis_challenges}
 \end{figure*}

In addition to the categorization described above, we took a deeper look into the results to understand the prevailing reports of challenges among our respondents and across different demographics. We avoid using the numerical prevalence of evidence to indicate the importance or criticality of any challenge. However, when presenting the results, we use supplementary and corroborative counting of the responses to triangulate the qualitative analysis~\cite{hannah2011counting}. The majority of respondents reported challenges related to \textsc{Work-Life Balance Issues} (54.7\%), \textsc{Glass Ceiling} (40.6\%), and \textsc{Sexism} (21.9\%). Figure \ref{fig:segmented_analysis_challenges} illustrates those three categories of challenges for each demographic.

We used the most representative demographic subgroup (see Table~\ref{tab:demographics}) to present the percentages, which reflect the number of participants who mentioned any difficulties that were classified under each category of challenge. Most of the respondents who reported challenges related to \textsc{work-life balance issues} are married and live in India, with proportions higher than 80\%. No demographic was so prevalent for glass ceiling and sexism, demonstrating that both challenges were reported by a broader subgroup.

We also calculated the odds ratio for each of the three most reported challenges, considering the demographics detailed in Section~\ref{sec:method:data_analysis}. Table~\ref{tab:odds} presents the results of the odds ratio for each category of challenge. According to our sample, women with more experience in the software industry have higher odds (3.62x) than those with less experience to report challenges related to \textsc{work-life balance issues}. The odds that married women report challenges related to \textsc{work-life balance issues} are 5.08x higher than for single or divorced women.

When analyzing the intersectionality of the demographics, we learned that all of the 20 women who have more than 10 years of experience in the software industry and reported challenges that were categorized into \textsc{work-life balance issues} are married or live in a domestic relationship with children. Regarding the \textsc{glass ceiling}, we also observed that more than the half of the women who reported this challenge have more than 10 years of experience in the software industry, are married or live in a domestic relationship, and live with children.

\MyBox{We found eight categories of challenges, with the most mentioned ones categorized into work-life balance issues, glass ceiling, and sexism. Women from our sample with more experience in the software industry and who were married were more likely to report challenges related to work-life balance issues.}

\subsection{RQ2: What are possible actions to mitigate the identified challenges, from the women's perspectives?}
\label{sec:results:rq2}

\begin{figure*}[hbt]
\centering
\includegraphics[width=1\textwidth]{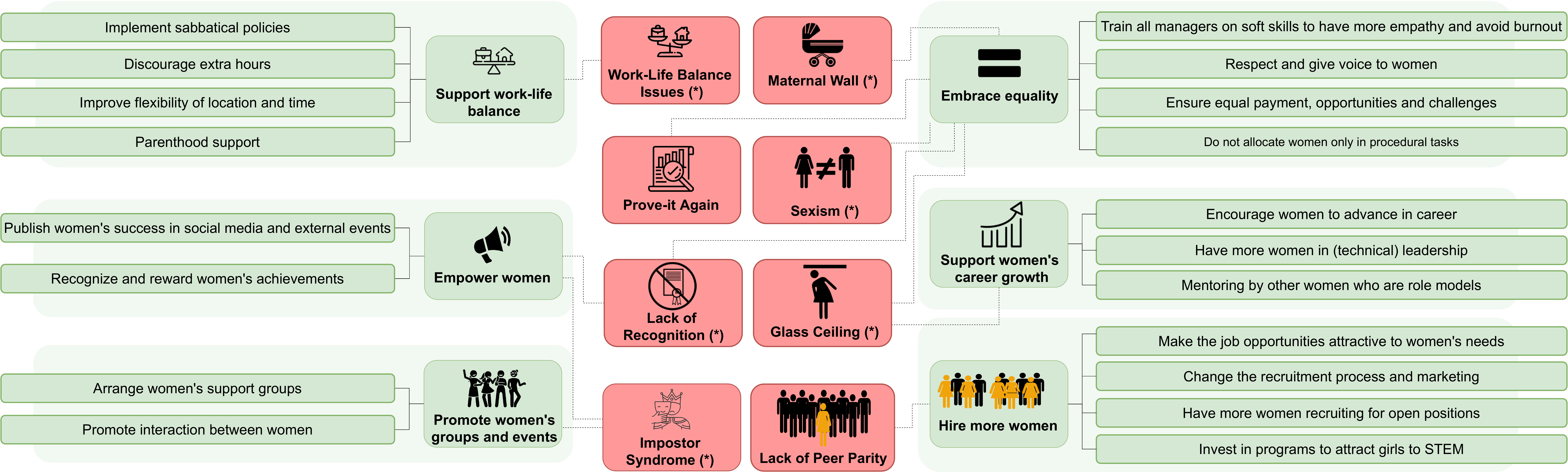}
\caption{Strategies suggested by women for the company to increase women in software development teams}
\label{fig:strategies}
\end{figure*}

We now present the actions recommended by the participants to help mitigate the challenges and retain more women in the company. Our analysis revealed six categories that explain the actions suggested by women, which we present in Fig.~\ref{fig:strategies}.


Table~\ref{tab:codes_suggestions} presents the number of participants whose responses fit in each category. Most of the respondents suggested the company \textsc{Hire More Women} (35.7\%), \textsc{Support Work-Life Balance} (30.0\%), and \textsc{Embrace Equality} (28.6\%). We used the most representative demographic subgroup (see Table~\ref{tab:demographics}) to present the percentages, which reflect the number of participants who mentioned any difficulties that were classified under each category of suggestion. In the following, we present our findings organized by category.

\begin{table*}[hbt]
\centering
\caption{Representative examples of answers to the suggestions' open question, number and percentage of women whose answer were coded for each category}
\label{tab:codes_suggestions}
\begin{tabular}{c|l|r|r}
\hline
\toprule
 \textbf{Suggestion} & \textbf{Representative examples} & \textbf{\#}  & \textbf{\% (n=70)}  \\
 \hline

\toprule

\begin{tabular}[c]{@{}c@{}}Embrace Equality\end{tabular} & \begin{tabular}[l]{@{}l@{}}\textit{`Speaking gender-diversity alone will not be suffice, it should also be reflected in the equal payouts}\\\textit{of deserving female candidates"} (S28) \end{tabular}& 20 & 28.6\% \\ \hline

\begin{tabular}[c]{@{}c@{}}Support Women's\\ Career Growth\end{tabular} & \begin{tabular}[l]{@{}l@{}}\textit{``Promoting women to senior jobs and leadership would help younger talents to identify }\\\textit{themselves with the company, giving them confidence and more prospects of continuing their}\\\textit{career in the company"} (S65)
\end{tabular}& 14 & 20.0\% \\ \hline
\begin{tabular}[c]{@{}c@{}}Hire More\\ Women\end{tabular} & \begin{tabular}[l]{@{}l@{}}\textit{``Active search for female talents"} (S39) \textit{Conduct women-only drives to hire fresh talent from girls}\\\textit{colleges."} (S59) \textit{``Publicize openings in workshops that are focused on women in market"} (S63)
\end{tabular}& 25 & 35.7\% \\ \hline
\begin{tabular}[c]{@{}c@{}}Promote Women's\\Groups and Events\\ Women\end{tabular} & \begin{tabular}[l]{@{}l@{}} \textit{``Virtual meetings, debates.."} (S66) \textit{``diversity events, so minorities to feel more valued."} (S40)
\end{tabular}& 3 & 4.3\% \\ \hline
\begin{tabular}[c]{@{}c@{}}Empower Women\end{tabular} & \begin{tabular}[l]{@{}l@{}} \textit{``Be more active on social media and do external open talks with woman from Ericsson talking} \\ \textit{about their work."} (S10) \textit{``Deserving women should be recognized and rewarded"} (S28) \\
\end{tabular}& 7 & 10.0\% \\ \hline

\begin{tabular}[c]{@{}c@{}}Support \\Work-Life Balance\end{tabular} & \begin{tabular}[l]{@{}l@{}}\textit{`Flexible work hours, and to focus on ensuring work-life balance. Workaholics tend to breach the}\\ \textit{latter, not just their own, but the team's too in a collaborative environment."} (S43)\\
\textit{`Enable women taking a break from career due to motherhood to return to the workforce. That is}\\ \textit{where we loose them."} (S2)\end{tabular} & 21 & 30.0\% \\ \hline

\bottomrule
\end{tabular}

\resizebox{0.75\textwidth}{!}{%
\tiny
\begin{tabular}{r}
The total per suggestion of improvement is not the sum of the respondents since the participants often provided an answer that was categorized into \textbf{more than one suggestion}.
\end{tabular}
}
\vspace{-0.2cm}

\end{table*}

\subsubsection{Embrace Equality}
Using the definition provided by UNESCO: ``Gender equality exists when all genders enjoy the same status and have equal conditions, treatment, and opportunities for realizing their full potential, human rights and for contributing to and benefiting from economic, social, cultural and political development.''~\cite{sharp2016unesco}. 

In terms of cultural improvements, the participants suggested the company \textsc{train all managers in soft skills to have more empathy and avoid burnout} in their teams, so they can also \textsc{respect and give voice to women}. 

In terms of process improvements, participants called for more  equality in terms of payment, opportunities, and challenges. Providing \textit{``equal payment between genders''} (S28), \textit{``equal opportunities without considering whether it is a women who is applying''} (S88), and \textit{``giving [women] equal challenges like [those] given to the male employees which allow them to venture into more of learning and become confident''} (S54). 

Finally, our participants asked to \textit{``\textsc{not allocate women only in procedural tasks}, but include [them] in projects in which they feel [they are a] part [of the team], responsible and that can challenge their skills''} (S62). This can give them equal opportunities and break the glass ceiling.

\subsubsection{Support Women's Career Growth}
To break the glass ceiling, the participants mentioned the need to \textsc{encourage women to advance in their career}, \textit{``moving ahead in other streams apart from people management roles as well and \textsc{having more women in technical leadership roles} where very few women seem to step in''} (S30). 

Mentoring can \textit{``help younger talents to identify themselves and give them confidence and more prospects of continuing their career in the company''} (S65). Prepare women to advance in career laddering by \textsc{mentoring by other women who are role models}, which can happen during \textit{``programs for women in leadership roles''}, \textit{``showing them how to grow, by teaching the skills that will help them get recognized in the crowd and make her feel valued''} (S42).

\subsubsection{Hire More Women}
To hire more women, our participants suggested changes to the job opportunities the company offers and to the recruitment process, and to invest in marketing \. Regarding the job opportunities, they suggest the company \textsc{make them attractive to women's needs}, create more part-time positions, and reserve positions prioritized for internal candidates and women. 

Regarding the action related to \textsc{change to the recruitment process}, besides \textsc{having more women as recruiters}, the suggestions included transparency about the required skills, advertising job openings to women's groups and events, and raising awareness about how the company supports women's growth. Finally, the suggestions included base-level actions like \textsc{investing in programs to attract girls to STEM}, and being \textit{``active in the social media that are accessed by the audience (e.g., Snap/Instagram)''} (S90) so they would be \textit{``securing more talent women to join''} (S38) and \textit{``from a fresher level''} (S58).

\subsubsection{Promote Women's Groups and Events}
Considering that many teams still have only a few women, our participants mentioned that the lack of parity can be mitigated by having\textsc{women's support groups} and events for \textsc{\textit{``interaction between women from different departments}''} (S81). The events were suggested to target not only women, but \textit{``to celebrate differences and diversity (gender, cultural, age), so that minorities feel more valued''} (S40).

\subsubsection{Empower Women}
Empowerment is a strong strategy to fight the impostor syndrome~\cite{clance1987imposter} and one of the strategies to foster it is recognition~\cite{manohar2001human}. The study participants recommended \textsc{recognition and achievements' rewards}. Moreover, since women's empowerment refers to enhancement of their power and position in society, some participants suggested that ``sharing the stories of [the company's] women can encourage and motivate women employees in achieving the same'' (S49). Morever, empowerment could be achieved by \textsc{\textit{``publishing women's success stories in social media''} (S47) and external events}.

\subsubsection{Support Work-Life Balance}
Sabbaticals are paid leaves for personal and professional development reasons~\cite{miller1998case} to promote well-being, and are also beneficial for the company, as they increase future productivity~\cite{davidson2010sabbatical}. The participants suggested to \textsc{implement sabbatical policies} so they can \textit{``take a break from [their] career''} (S2) to rest, dedicate some time to their family, and acquire new knowledge. 

Another suggestion was to \textit{\textsc{``discourage extra hours}, as the workaholics tend to breach the latter, not just their own, but the team's too in a collaborative environment''} (S43). 


Maternity leaves usually relate to each country's law. Some countries mandate a long maternity leave, while others mandate a shorter one~\cite{givati2012law}. The participants protested that ``maternity leaves should not be considered an impediment for a woman to grow'' (S54) and asked for the company to \textsc{assist maternity} by implementing own rules to extend the paid leave to 1 year, even \textit{``beyond local laws of the country''} (S22). 


%
 \MyBox{Participants called for actions to embrace equality and help reduce the sexist culture inside the company, and to mitigate the biases that create maternity wall and prove-it-again effects. Initiatives to support women's growth were suggested to start breaking the glass ceiling. Hiring more women and organizing women's groups could foster peer parity, and women's empowerment could reduce impostor syndrome and provide recognition. Sabbaticals and discouraging long work hours were mentioned as ways to improve the work-life balance.}

\section{Discussion}
\label{sec:discussion}

In this section, we present a more in-depth discussion of our results in the context of the literature.

\noindent\textbf{Work-Life Balance Issues faced by mothers.} Work and family are the two most important domains in a person's life and their interface has been the object of study for researchers worldwide~\cite{sturges2004working}. As women assume the role of working professional in addition to their traditional role of homemaker, they are under great pressure to balance their work and personal lives~\cite{sturges2004working}. The societal role expectations, women's career ambitions, and the nature of the IT industry challenges the way women manage their professional and personal lives~\cite{haq2013intersectionality}. 
The COVID-19 pandemic and the need to work from home cast new light on these issues. While it brought more flexibility to many workers (which is the case of the company studied), it also brought new challenges~\cite{ford2020tale,ralph2020pandemic}. For a great share of the population, it became hard to separate personal and professional life. Women felt this more than men, given the aforementioned societal expectations\cite{machado2020gendered}.

Work-life balance is a challenge that happens in and beyond the IT industry. In Japan, work-life balance is a general challenge, and the low numbers of women in medicine reflect the societal belief that careers and motherhood do not mix~\cite{ramakrishnan2014women}. In contrast, Scandinavia has similar numbers of men and women physicians, which has coincided with the emergence of progressive work–life policies, the belief that women can combine motherhood and employment, and changing expectations of work-life balance. This shows that it is possible and that the mindset in the software industry needs to change.  

The lack of parent support is one of the major challenges that Indian companies within the IT sector have been trying to navigate over the past decade to ensure a gender inclusive workplace~\cite{raghuram2017women}. This is reinforced by our work,in which women provided suggestions on how to \textsc{improve parental policies to support families} and mitigate this challenge, as we showed is Section~\ref{sec:results:rq2}. Sponsoring child care, preferentially in the office and specially for young children, providing adequate maternity leave beyond the relevant country's laws, and also providing more flexibility in work hours and location were some of the suggestions for the company to take in order to mitigate the challenges of work-life balance issues related to motherhood. This is harder to implement in many cases due to the countries' legislation and the local culture, which influence how organizational policies are defined.

\textbf{Prepare women to break the glass ceiling.} Analogous to the IT industry, women's barriers in the medical profession and their ability to rise to leadership positions are also influenced by social and cultural context~\cite{ramakrishnan2014women}. Similar to software teams, where women are instrumental to reducing community smells~\cite{catolino2019gender}, in international relations the collaboration between women delegates and women civil society groups positively impacts and brings more durable peace when negotiating peace agreements~\cite{krause2018women}. By analyzing the career trajectories of women executives across a variety of sectors, Glass and Cook~\cite{glass2016leading} concluded that while attaining promotion to leadership is not easy, serving in a high position can be even more challenging. Although women can be more likely than men to be empowered to high-risk leadership positions, they often lack the support or authority to accomplish their strategic goals. As a result, women leaders often experience shorter tenures compared to their peers who are men~\cite{glass2016leading}. 

The respondents provided suggestions to \textsc{support their career growth} to mitigate this challenge, including having women to inspire, encourage and mentor other women, as we showed is Section~\ref{sec:results:rq2}. Providing training for mentors on topics such as speaking up on behalf of women who are being disrespected in meetings, managing bias in the workplace, and raising awareness of micro-aggressions at work are some examples of what should be included as part of standard training and preparation for mentors~\cite{giscombe2017creating}. The mentoring program can start by having a different woman leader each month discuss her career trajectory and the benefits and challenges of holding her job. Women could share their techniques for managing time, balancing family and career demands, making themselves heard by men, and highlighting how they learn new skills on a regular basis~\cite{yen2005advance}. Besides joining ongoing support groups, women can be assigned to formal mentors for one-on-one regular meetings~\cite{kosoko2006mentoring}.


\textbf{Combining synergistic suggestions.} One option for companies looking to improve women's participation is to combine strategies that are synergistic. The company can start by implementing simple, but structured actions combining ideas from more than one strategy. For example, by publishing success stories of women on media, the company can \textsc{empower women} and also attract and \textsc{hire more women}. Another action that combines synergistic strategies is to \textsc{arrange women-only groups} and analyze messages to implement feasible changes to problems that are being actively discussed and could potentially cause women to leave, as when women report facing hostile sexism. The literature reports that women experience computing environments differently due to sexism and racism, both historically and as part of the current culture~\cite{barker2009exploring,margolis2002unlocking,margolis2017stuck,strayhorn2012exploring}, potentially leading them to feel unwelcome and lacking of sense of belonging~\cite{sax2018sense}, and ultimately to leave~\cite{espinosa2011pipelines}.

\textbf{Some problems come from beyond the company gates.} Even in a company like Ericsson, which cares about Diversity \& Inclusion, ``untying the mooring ropes" of socio-cultural problems is difficult. Historically, the social differences influenced by gender roles (i.e., the roles that men and women are expected to occupy based on their sex) may be amplified because of the gendered division of housework and child care tasks, especially for mothers of young children. 
Impostor syndrome~\cite{wolff2020prevents,Lee.Carver:2019,balali2018}, Sexism~\cite{Canedo2021SBES,singh2019women2,calvo2021visible,Nafus2012NMS}, Lack of Peer Parity~\cite{powell2010gender,canedo2020work}, Prove-it-Again~\cite{imtiaz2018}, Glass Ceiling~\cite{Canedo2021SBES} and Work-Life Balance issues~\cite{machado2020gendered,Lee.Carver:2019} were challenges reported by women from the present study and also reported by both women in other software development contexts and in F/OSS. Some problems surpass the organization and are related to the local culture of the employees and managers. There are problems that go beyond the company's gates and bump into the society, which many times contributes to this cultural legacy. One example is the ``trailing spouse", when a person who follows his or her life partner to another city because of a work assignment~\cite{harvey1998dual,branden2014gender}. Moreover, during the COVID-19 pandemic, a longer ``double-shift'' of paid and unpaid work in a context of school closures and limited availability of care services have contributed to an overall increase in stress, anxiety around job insecurity, and difficulty in maintaining work-life balance among women with children~\cite{globalgendergap2021}. However, there is also space for improvement in the organization, and Ericsson is committed to implementing the suggested changes to mitigate the challenges faced by its women employees.


\subsection{Implications to the company}
We presented the results to the managers of the studied suborganization and to the managers of Human Resources department. The feedback was very positive regarding the usefulness of the research. Managers considered the results helpful for the company to understand the current situation and to decide about actions can mitigate the challenges that women currently face and avoid them leaving the company. For the suggestions that are already in place, such as publishing successful stories of women and support groups, the company plans to expand and raise awareness to the employees. \textsc{Sexism} is considered an unacceptable behavior to Ericsson. The managers of the studied suborganization already started to have collective meetings with the team to spread the message and remind that sexism is not tolerated by the company. Following, the company is planning recurrent meetings to plan  strategies that address the suggestions provided by the participants. In addition, Ericsson plans to raise awareness about the solutions that already exist (and maybe women are not aware of) and additional actions that can complement the suggestions to mitigate the reported challenges.

\section{Related Work}
\label{sec:related_work}

Although diversity is a multidimensional concept that refers to the variety of representations that exist within a group~\cite{williams1998reilly,albusays2021diversity}, gender (with a focus on women), is the most explored aspect of diversity in software engineering literature~\cite{menezes2018diversity,silveira2019systematic}. The prevalence of gender as the most studied diversity aspect can be explained by the fact that the technology professions are known to be male dominated, despite the fact that programming was originally seen as a female occupation~\cite{ensmenger2012computer, hicks2017programmed, bjorn2019femtech, bjorn2021intertextual}. According to a longitudinal study that evaluated 50 years of data and the evolution of code contributions since 1970, \citet{zacchiroli2020} showed that woman developers’ contributions remain low when compared to those of men. Although the study found that men have always authored more open source code than women, the gap has begun to narrow and women are slowly gaining space.

The biases and challenges faced by women in software development teams have been investigated in different industry cases~\cite{Canedo2021SBES, wolff2020prevents, musungwini2020challenges,orser2012perceived}. The Finish women from \citet{wolff2020prevents}'s study reported lack of self-efficacy, which is a possible predecessor of \textsc{impostor syndrome}, also found in our study. Also similar to our results, the Brazilian women from \citet{Canedo2021SBES}'s study also reported \textsc{hostile sexism}, including discrimination and bias, and \textsc{benevolent sexism}, when women do not receive the more complex tasks, and \textsc{glass ceiling}, as only few women perform a leadership role in their team. Another study with  Brazilian women during the COVID-19 pandemic~\cite{machado2020gendered} revealed that women faced even more \textsc{work-life balance issues}, lacking support with housework and child care responsibilities. As opposed to the \textsc{glass ceiling} reported by \citet{Canedo2021SBES}'s study and our results, the main challenge reported by the Zimbabwean women from \cite{musungwini2020challenges}'s study was lack of digital exposure and career guidance from young age. So far, the studies point at generic aspects, focusing on understanding a broader, more comprehensive picture of the challenges faced by women. In our work, we evolve the findings by diagnosing an organization that is investing in programs to attract and retain women. 

Women who contribute to F/OSS projects have reported some of the challenges also reported by women from the present study: work-life balance issues~\cite{Lee.Carver:2019}, impostor syndrome~\cite{Lee.Carver:2019,balali2018}, lack of peer parity~\cite{powell2010gender,canedo2020work}, prove-it-again~\cite{imtiaz2018} and sexism~\cite{singh2019women2,calvo2021visible,Nafus2012NMS}. Non-inclusive communication is faced by F/OSS women in code reviews and mailing lists~\cite{paul2019expressions,balali2018,qiu2019signals}, and was reported as an aspect of the hostile sexism faced by women in the present study, which they described taking place during meetings. While in F/OSS women face bias against contributions when they explicitly identify as women\cite{terrell2017gender,canedo2020work}, in our study we found a lack of recognition. Stereotypes manifest the common expectations about members of certain social groups. Both the descriptive (how women are) and prescriptive (how women should be) gender stereotypes and the expectations they produce can compromise a woman's career progress~\cite{heilman2001description, heilman2012gender}. In F/OSS projects, women reported facing stereotyping that box them into specializations despite their manifest protest~\cite{vedres2019gendered} and being treated by men as if they were their mothers, asking for advice about how to dress and behave but refusing to enter into a technical dialogue~\cite{Nafus2012NMS}. While women in our study reported also facing stereotypes, for them it was part of the \textsc{Maternity Wall} and receiving fewer responsibilities because they had children. And whereas F/OSS women reported that they faced obstacles to finding a mentor, since upon discovering their mentee's gender, men mentors can treat the relationship as a dating opportunity~\cite{Nafus2012NMS}, this was not reported by women from the present study.

\section{Threats to Validity}
\label{sec:threats}

There are some limitations related to our research results.

\textbf{Internal validity}. The characteristics of our sample may have influenced our results. Although the company has office in several locations, a great part of the responses (42 out of 64) were from women who live in India. Thus, the challenges and suggestions can reflect some of the specific socio-cultural problems and aspirations for women from this country. The most prevalent nationality of employees overall from the studied suborganization is Indian (55\%).

\textbf{External Validity}. The results are valid for the studied suborganization of Ericsson and additional research is necessary to investigate the challenges and suggestions in other contexts.

\textbf{Survival bias}. Our results reflect the opinion of current employees. Therefore, to increase women's participation by fully understanding the reasons they might leave, we acknowledge that additional research is necessary to understand the point of view of the women who left the company. To mitigate this threat, we asked the participants about reasons that prompted a woman they know to leave the company.

\textbf{Recall bias.} As our questions were open-ended, our results could be impacted by either salience bias, where respondents focus on definitions that are prominent or emotionally striking and not necessarily all the factors that matter; or by memory bias, where participants answered questions based on what they can first recall. However, topics that are relevant to the respondent often emerge from the spontaneous answers. 

\textbf{Data Consistency}. Consistency refers to ensuring that the results consistently follow from the data and there is no inference that cannot be supported after data analysis~\cite{MerriamBook}. The group of researchers performed the qualitative analysis of questionnaire's responses. We had weekly meetings to discuss and adjust codes and categories until reaching agreement. In the meetings, we also checked the consistency of our interpretations. All analysis was thoroughly grounded in the data collected and exhaustively discussed amongst the whole team. The team includes researchers with extensive experience in qualitative methods.

\textbf{Theoretical saturation}. A potential limitation in qualitative studies regards reaching theoretical saturation. From participants in this study with different backgrounds and perceptions about the studied phenomenon, we received 64 responses for the challenges question and 70 for the strategies. The participants were diverse in terms of experience, tenure, age, family status. Therefore, although theoretical saturation cannot be claimed, we believe that we obtained a consistent and comprehensive account of the phenomenon for the studied case. After analyzing the $40^{th}$ response of challenge and the $29^{th}$ response of suggestion we did not find any new categories, using the existing categories for the following 24 challenges and 41 suggestions.

\section{Conclusion}
\label{sec:conclusion}

This paper presents a case study aiming at understanding the challenges faced by women in a large software company and collecting strategies to increase the number of women. We found that even with the commitment with diversity and inclusion from Ericsson, women still perceive challenges and call for changes. 


We also showed that the cultural structural sexism present in society is mirrored in the professional environment. There is still a long work ahead for Ericsson, for the software industry, and for us, as society, to create a more diverse and inclusive environment. We hope our results will enlighten actions towards reducing the perceived challenges and (more importantly, maybe) increasing awareness about the structural and cultural hurdles imposed on women that negatively influence diversity in the software industry.


\bibliographystyle{ACM-Reference-Format}
\bibliography{sample-base}

\end{document}